\documentclass[runningheads]{llncs}

\usepackage{amsmath}
\usepackage{tikz}
\usepackage{xspace}
\usepackage{subcaption}
\usepackage{cleveref}

\title{The PoW Landscape in the \\ Aftermath of The Merge}
\author{Lucianna Kiffer \and Sophia Skorik \and Yann Vonlanthen \and Roger Wattenhofer}

\institute{ETH Zurich, Switzerland \\
    \email{\{lkiffer,yvonlanthen,wattenhofer\}@ethz.ch, \\ sophia.skorik@inf.ethz.ch}}
\authorrunning{L. Kiffer et al.}

\begin{document}
\maketitle
\vspace{-0.3in}
\begin{abstract} 
    On 15th September 2022, The Merge marked the Ethereum network's transition from computation-hardness-based consensus (proof-of-work) to a committee-based consensus mechanism (proof-of-stake). As a result, all the specialized hardware and GPUs that were being used by miners ceased to be profitable in the main Ethereum network. Miners were then left with the decision of how to re-purpose their hardware. One such choice was to try and make a profit mining another existing PoW system. In this study, we explore this choice by analyzing the hashrate increase in the top PoW networks following the merge. Our findings reveal that the peak increase in hashrate to other PoW networks following The Merge represents an adoption of at least 41\% of the hashrate that was present in Ethereum, with 12\% remaining more than 5 months later. Though we measure a drastic decrease in profitability by almost an order of magnitude, the continued presence of miners halts claims that power consumption was instantly addressed by Ethereum's switch to PoS.
\end{abstract}

\section{Introduction}
\label{sec:intro}

Proof-of-work (PoW) was first introduced as a way to prevent spam emails and denial-of-service attacks \cite{dwork1992pricing,back2002hashcash}. By requiring users to spend resources solving computationally hard cryptographic challenges, it is possible to put a price on bot and spam activity.
Soon after, the first applications making use of PoW for digital money were introduced by Hal Finney \cite{finney2004rpow} and Nick Szabo \cite{szabo2005bit}. Ultimately, circumventing the need for a central entity using PoW, Satoshi Nakamoto built the first permissionless digital currency \cite{nakamoto2009bitcoin}. Bitcoin's meteoric rise starting in 2008 truly brought PoW into the spotlight and made mining a highly lucrative venture \cite{taylor2017evolution}.

Thus began a never-ending race of producing the most computation with the fewest resources \cite{han2019demystifying}. As a consequence, many PoW algorithms quickly became profitable to mine only using application-specific integrated circuits (ASICs) \cite{taylor2017evolution}. At the same time, numerous cryptocurrencies emerged, many introducing their flavors of PoW \cite{litecoin,kalodner2015namecoin,monero,hopwood2016zcash}. Some of which designed algorithms to be resistant to ASICs \cite{chepurnoy2019autolykos} in a bid to lower the barrier to entry and improve democratization and decentralization \cite{han2019demystifying}. These currencies make use of memory-hard mining algorithms that can be most efficiently run on consumer-grade graphical processing units (GPUs) \cite{meneghetti2020surveypow,han2019demystifying}.

The rise of Ethereum \cite{buterin2013ethereum} starting in 2015 stands out, not just for its memory-hard PoW algorithm called Ethash, but also for allowing for not just money transfer but universal computing across untrusted participants. 
Its rapidly growing popularity was felt throughout the entire GPU industry \cite{toms2020worsen}, as Ethereum miners alone are estimated to have bought at least 10\% of all produced GPUs between October of 2020 and March of 2022 \cite{toms2022spent}. Spiking prices, longer wait times, and chronically low supply have been the result for years \cite{economist2021blame}.

Amidst growing concerns regarding PoW's tremendous energy consumption \cite{wendl2023environmental} in the advent of the climate crisis, Ethereum considered PoW as a temporary solution, with the goal to move its Sybil resistance mechanism to proof-of-stake (PoS) \cite{buterin2016pos}. PoS prevents Sybils, by giving the voting power to the system's stakeholders \cite{king2012ppcoin}. After years of development and numerous delays, Ethereum finally switched to PoS on 15th September 2022, an event referred to as The Merge \cite{ethereum2022themerge}. 

In this work, we explore the consequences of this long-anticipated switch. To this end, we perform a longitudinal study to quantitatively measure how much GPU mining power remains in the space after The Merge, the profitability of the ongoing mining business, as well as miner distribution. Based on this data, we analyze the impact of The Merge on energy expenditure, the decentralization of other currencies, as well as the consequences for miners, mining pools, users, and manufacturers. Our \textbf{contributions} are the following:
\begin{enumerate}
    \item We collect an extensive dataset containing blockchain, market, and miner data for the main currencies employing memory-hard PoW. By additionally scraping GPU performance data we homogenize the different PoW algorithms to allow their comparison.
    \item We measure that at least 12\% of Ethereum's mining power remained active months after The Merge even though mining profitability has drastically reduced by almost an order of magnitude (87.7\%). This provides a more nuanced perspective to claims which say The Merge reduced energy consumption by 99.95\% instantaneously \cite{ethereum2022themerge}. 
    \item We explore which mining pools remained active across the different emerging forks such as Ethereum PoW \cite{ethw} and Ethereum Fair \cite{ethf}, as well as other currencies. We show that surprisingly, in most cases, decentralization wasn't negatively impacted.
    \item Finally, we discuss the effects of The Merge on the major stakeholders, such as miners and GPU manufacturers.
\end{enumerate}

\section{Data Collection}
\label{sec:datacollection}
    \begin{table}[t]
            \centering
            \tiny
            \caption{Computation of ethash factor based on GPU hashrates}
            \label{tab:ethash-factors}
            \begin{tabular}{|l|c|c|c|c|c|c|c|c|c|c|c|c|c|}
            \hline
                Algorithm & Nvidia &  Nvidia &  Nvidia & Nvidia & Nvidia & Nvidia & AMD & AMD   & AMD    & AMD   & AMD    & AMD & \\
                          & GTX    &  RTX    & RTX     & RTX    & GTX    & RTX    & RX & Radeon & RX     & RX    & RX     & RX & median  \\
                          & 1080   & 3090    & 2080    & A5000  & 1080Ti & 3060Ti & 6800& VII   & 5700XT & 6600XT& 6900XT & 480& \\
                \hline
                EtcHash & 1.01&1&1&1&\textbf{1.14}&1.01&1&1&1&1&1&1&1 \\
                Autolykos &1.72&2.27&1.95&2.20&2.16&\textbf{2.45}&1.83&2.23&1.81&1.82&1.81&1.94&1.94 \\
                KawPow & 0.46&0.41&\textbf{0.69}&0.41&0.56&0.50&0.52&0.37&0.45&0.50&0.52&0.41&0.48 \\
                
                \hline
            \end{tabular}
        \end{table}

    Our first goal is to determine which cryptocurrencies absorbed the bulk of the Ethereum mining power or \textit{hashrate}. To do this we first look up the top 100 by marketcap \cite{coinmarketcap} PoW currencies' hashrates on a popular miner website \cite{2miners} on the days surrounding the merge. We make a list of all that had a \textit{visually noticeable increase} in their system hashrates (i.e., what we see in Figure~\ref{fig:bigplot}) as we want to distinguish those likely directly impacted by The Merge, from other ongoing trends\footnote{Note in Table~\ref{tab:hashrate-increas-ethash} increases of 2x to 5x their relative hashrates.}. In doing so, it is possible we miss some hashrate that was less noticeably absorbed by other systems. We also include two new forks that split from Ethereum following The Merge; Ethereum Fair \cite{ethf} and Ethereum PoW \cite{ethw}. From these initial candidates, we want to gather which networks absorbed the majority of the Ethash computing power. Since these blockchains run different mining algorithms, we need to convert each network's hashrate to their Ethash equivalent.
    
    \vspace{-0.2cm}
    \paragraph{Ethash Conversion Factors.}
        The hashrates generated by different GPUs for a specific algorithm may vary due to differences in their design and processing capabilities. To account for this, we take 12 of the top Nvidia and AMD GPUs recommended for Ethereum mining in 2022 \cite{bestgpus1,bestgpus2,bestgpus3}. For each GPU we determine the hashrate per algorithm and calculate its factor in relation to Ethash hashrate\footnote{Though the use of ASICs was not prominent for Ethash, we do conservatively capture their hashing power as ASICs would suit only Ethereum forks at factor 1.}. Since different mining sites report slightly different values for a given GPU-algorithm combination, we gather multiple values from various sources and take the median factor across them \cite{minerstat,whattomine,2cryptocalc,minerbay}.

\begin{table}[t]
            \centering
            \caption{Hashrate Increase for top 20 GPU-minable coins affected by The Merge.}
            \label{tab:hashrate-increas-ethash}
            \begin{tabular}{|c|l|c|c|r|r|}
                \hline
                \textbf{}      & \textbf{}    & \textbf{Network} & \textbf{Network} & \textbf{Median~~}& \textbf{Minimum} \\ 
                \textbf{Coin} & \textbf{Algorithm}  & \textbf{Hashrate} & \textbf{Hashrate}&\textbf{Increase~} &  \textbf{Increase~~}\\
                & &(7th Sep) & (15th Sep)& \tiny{(Ethash TH/s)}& \tiny{(Ethash TH/s)} \\  \hline\hline
                Ethereum & Ethash & 873.930 TH/s & - & - & -                                    \\ \hline
                EthereumClassic   & EtcHash  & 49.67 TH/s & 307.99 TH/s  & 258.32 & 226.60      \\ \hline
                Ergo    & Autolykos          & 26.13 TH/s & 149.39 TH/s  & 63.53  &  50.31      \\ \hline
                EthereumPOW & Ethash              & - &  46.16 TH/s   & 46.16    &  46.16                   \\ \hline
                Ravencoin & KawPow           & 3.83 TH/s  & 18.82 TH/s   & 31.23  &  21.72      \\ \hline
                Ethereum Fair & Ethash       & - &  5.93 TH/s    & 5.93    &   5.93                   \\ \hline
                Neoxa   & KawPow             & 1.91 TH/s & 3.75 TH/s  & 3.83    &  2.67         \\ \hline
                Conflux  & Octopus           & .84 TH/s & 2.53 TH/s   &  4.46  &   1.32         \\ \hline
                Kaspa   & kHeavyHash     & 75.20 TH/s & 95.22 TH/s    &  2.21  &  1.32          \\ \hline
                Firo  & FiroPow          & 91.86 GH/s & 435.00 GH/s  &   .76   & .59            \\ \hline
                EtherGem  & Ethash       & 2.25 GH/s & 448.78 GH/s    & .45 &   .45             \\ \hline
                Sero   & ProgPow         & 49.82 GH/s & 182.88 GH/s  &  .33   &  .25            \\ \hline
                Expanse  & Ethash        & 25.63 GH/s & 264.73 GH/s  & .24    &  .24            \\ \hline
                Nimiq   & Argon2d-NIM    & 1.90 GH/s & 3.97 GH/s      &    .21 &   .21          \\ \hline 
                Vertcoin  & Verthash     & 1.54 GH/s & 3.57 GH/s  &     .20    &    .10         \\ \hline 
                Etho   & Ethash          & 4.95 GH/s & 84.45 GH/s & .08        & .08            \\ \hline
                Ubiq   & Ubqhash         & 17.45 GH/s & 83.06 GH/s  &   .07   &  .07            \\ \hline
                Quarkchain  & Ethash     & 24.71 GH/s & 82.76 GH/s  & .06    &   .06            \\ \hline
                Callisto  & Ethash       & 71.38 GH/s & 102.22 GH/s  & .03      & .03           \\ \hline
                Zano  & ProgPowZ         & 6.89 GH/s & 16.77 GH/s   & .03    &    .02           \\ \hline
                \hline
                \textbf{Total} & && & \textbf{418.13} & \textbf{358.13}
            \\ \hline
            \end{tabular}
\end{table}
        
        We take the median and maximum of these values to get an Ethash factor to convert each mining algorithm hashrate to Ethashes. We show these values for Etchash \cite{etc2023github}, Autokylos \cite{chepurnoy2019autolykos}, and KawPow \cite{fenton2018ravencoin} in Table~\ref{tab:ethash-factors}, with the maximum factors in bold. Due to some variance in values, we use the maximum factor to give us a lower bound of the Ethash in each system, assuming the optimal GPUs went to the respective systems. Using this conversion factor, we get a list of the top 20 coins whose hashrates were impacted by The Merge (cf. Table~\ref{tab:hashrate-increas-ethash}). We see that a peak of a minimum of 41\% (median of 48\%) of the Ethash from Ethereum went into the top 20 GPU-minable PoW coins. Note that other top PoW coins like Bitcoin, Dogecoin, and Litecoin do not show up on our list likely as they are primarily profitable to mine via ASICs mining and thus did not have noticeable gains in hashrate at the time of The Merge.  We see that Ethereum Classic, Ergo, EthereumPoW, Ravencoin, and Ethereum Fair jointly absorbed 98\% (97\%) of the peak of the minimum (median) noticeable Ethash hashrate increase. Thus we choose to focus on these systems for the remainder of our analysis.

    \paragraph{Block data.} 
        In lieu of running clients in all 6 systems to collect blockchain data, we rely on block explorers \cite{oklink,ergoplatform,rvnexplorer} 
        to get all blocks of each system for the days surrounding The Merge and several months following. In total, we scrape block data from 1st September 2022 to 1st March 2023. In the block data, we get the block height, timestamp, miner ID, block reward plus transaction fees, and block difficulty. We also collect mining pool labels (if a miner address is associated with a particular mining pool) by cross-referencing additional block explorers \cite{etherscan,rvncoin}.
        
    \vspace{-0.2cm}
    \paragraph{Market data.} We scrape hourly price data for each coin from CoinGecko \cite{coingecko}.
\section{Analysis}
\label{sec:analysis}

\subsection{Mining Power Redistribution}
\label{sec:analysis:hashrate}

We begin by recomputing the network hashrates for each of the systems. The mining \textit{difficulty} parameter in each block is set such that blocks are found in \textit{expectation} at a rate determined by the \textit{block rate}. Dividing the difficulty by the expected block rate gives us the \textit{expected network hashrate} of the system, this is generally what popular mining statistics sites plot \cite{2miners}. Usually, the \textit{actual effective hashrate} of the system matches closely with the expectation, however, during times of sudden increase/decrease in hashrate, this is not the case. To capture such fluctuations, we divide the difficulty by the \textit{actual block rate}.

Due to the stochastic nature of mining, we additionally take the running average of the difficulty over the previous hour (3600 seconds) such that the computed hashrate $H$ at time $t$ is 
\begin{equation*}
    H_{network}(t) = \sum_{b \in blocks(t-3600,t)}{\text{diff}(b)}/3600 * \text{factor}_{network} 
\end{equation*}

\begin{figure}[t]
    \centering
    \includegraphics[width=\linewidth]{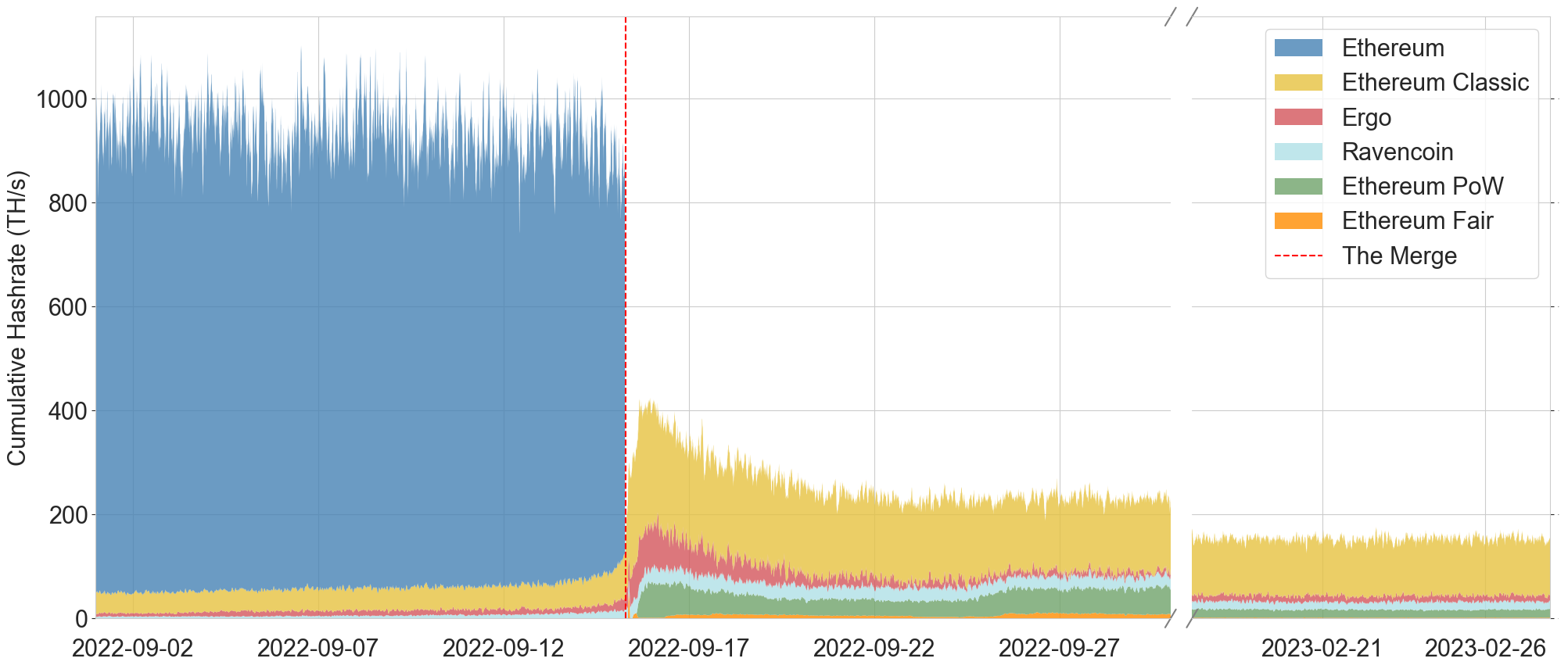}
    \caption{Cumulative minimum Ethash equivalent hashrate in each of the systems.} 
    \label{fig:bigplot}
    \vspace{-0.3cm}
\end{figure}

We use diff(b) to denote the difficulty of a block $b$, and $\text{factor}_{network}$ the Ethash conversion factor computed in Section \ref{sec:datacollection}, for the remainder of the paper we use the maximum factor. We plot the cumulative hashrate across the studied systems in Figure~\ref{fig:bigplot}, note a break in the x-axis during a period of steady decrease.

Prior to The Merge, each system had a relatively stable hashrate, Ethereum's hashrate being just 18.6\% from its all-time high a few months earlier \cite{2cryptocalc}. The end of Ethereum's mining meant a sudden and brief spike in the hashrate of all networks, reaching $41\%$ of Ethereum's hashrate. Within a few days, this spike leveled off and remained stable with a gradual decrease over the following five months. By the end of our measurement period, these five networks collectively still hold over 3x their pre-Merge hashrate, this increase is 30\% of the peak of the 15th of September and accounts for a minimum of 12\% (median of 14\%) of the pre-Merge Ethereum hashrate still remaining in these systems. In other words, at most 88\% of Ethereum's hashrate has left the ecosystem.

\subsection{Miner Distribution}

\begin{figure}[h!]
            \centering
            \begin{subfigure}{.496\textwidth}
                \centering
                \includegraphics[width=\linewidth]{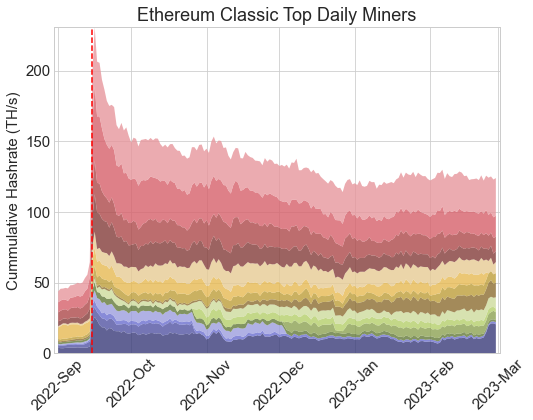}
            \end{subfigure}
            \begin{subfigure}{.48\textwidth}
                \centering
                \includegraphics[width=\linewidth]{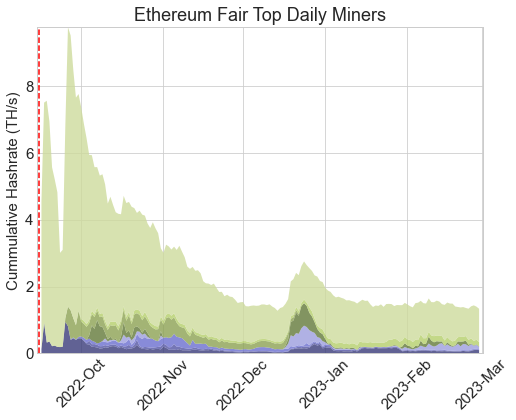}
            \end{subfigure}\\
            \begin{subfigure}{.496\textwidth}
                \centering
                \includegraphics[width=\linewidth]{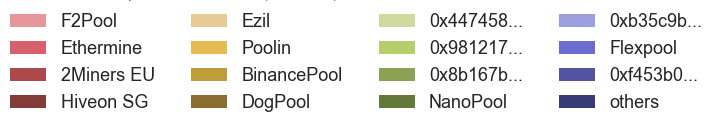}
            \end{subfigure}
            \begin{subfigure}{.45\textwidth}
                \centering
                \includegraphics[width=\linewidth]{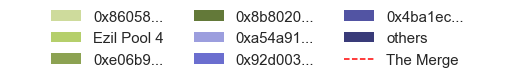}
            \end{subfigure}\\
            \caption{Mining Pool Distributions for Ethereum Classic and Ethereum Fair.} 
            \label{fig:topminers}
\end{figure}

\begin{figure}[h!]
            \centering
            \begin{subfigure}{.496\textwidth}
                \centering
                \includegraphics[width=\linewidth]{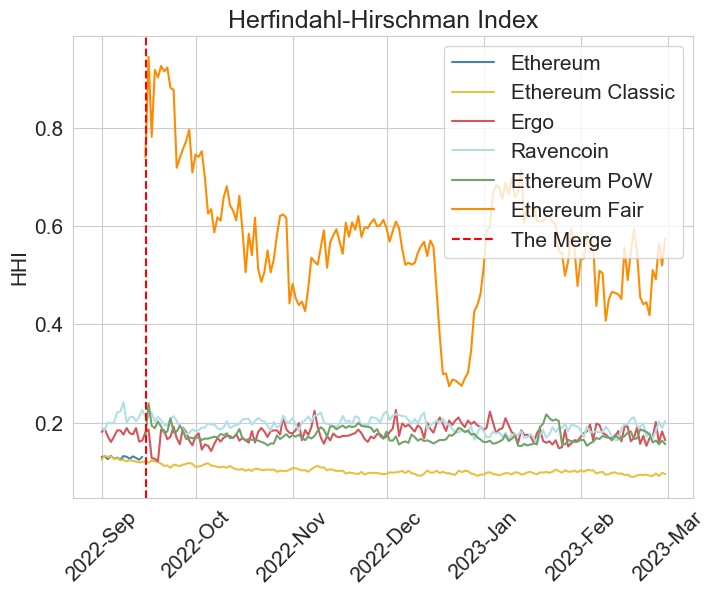}
            \end{subfigure}
            \begin{subfigure}{.496\textwidth}
                 \centering
                 \includegraphics[width=\linewidth]{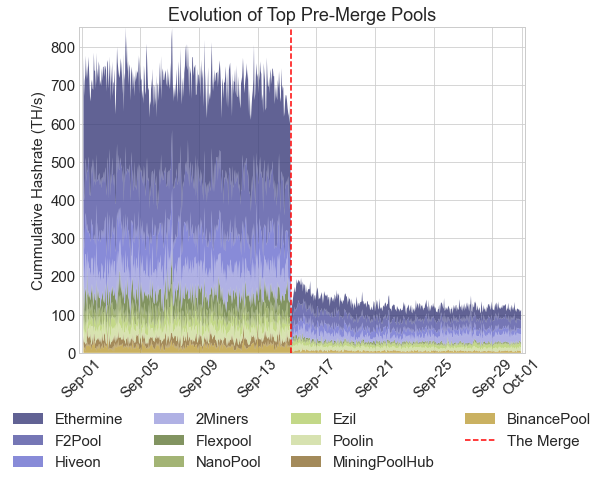}
             \end{subfigure}
            \caption{HHI for mining pool share distributions (left) and the cumulative hashrates of the top pre-Merge mining pools before and after The Merge (right).} 
            \label{fig:hhi_pools}
\end{figure}

We group the hashrate data by the top miner addresses in each system, including those addresses for which we have labeled as belonging to a mining pool. We plot the distributions in miner hashrate for Ethereum Classic and Ethereum Fair in Figure~\ref{fig:topminers}. We observe that most systems look very similar to what we see with Ethereum Classic, where most mining pools keep operating post-Merge with a growth in hashrate. Ethereum Fair on the other hand, from the onset, is under the control of a single large miner. We compute the Herfindahl-Hirschman Index for mining pool block share for each system in Figure~\ref{fig:hhi_pools} and further see that The Merge has little to no impact on the concentration of miners.

By taking common miner labels in all networks, we are also able to track a large portion of the pools following The Merge. Figure \ref{fig:hhi_pools} plots the cummulative hashrate of the largest mining pools over all systems before The Merge. We believe that for monetary reasons (switching between most profitable currencies) and for security reasons \cite{chatzigiannis2022diversification} most pools were already active across all platforms. Surprisingly, they make up a much lower portion of the total hashrate after the Merge. This could indicate more well-thought-out strategies from these pools.

\subsection{Mining Profitability}

We measure the block rewards and transaction fees that were collected by miners. Using the ensuing matching market price of a network's coin and the Ethash equivalent hashrate $H(t)$, we obtain the profitability $P(t)$, a metric describing the total US-dollar rewards per Ethash of computation done. \footnote{In practice, profitability can also be impacted by block stale, reject, or orphan rates and associated rewards or penalties. We do not consider these differences.}

\begin{equation*}
    P_{network}(t) = \frac{\left(\sum_{b \in blocks(t-3600,t)}\text{reward}(b)+\text{fees}(b)\right) * \text{price}_{network}(t)}{H_{network}(t) * 3600}
\end{equation*} 

\begin{figure}[t]
    \centering
    \begin{subfigure}{1.\textwidth}
        \centering
        \includegraphics[width=\linewidth]{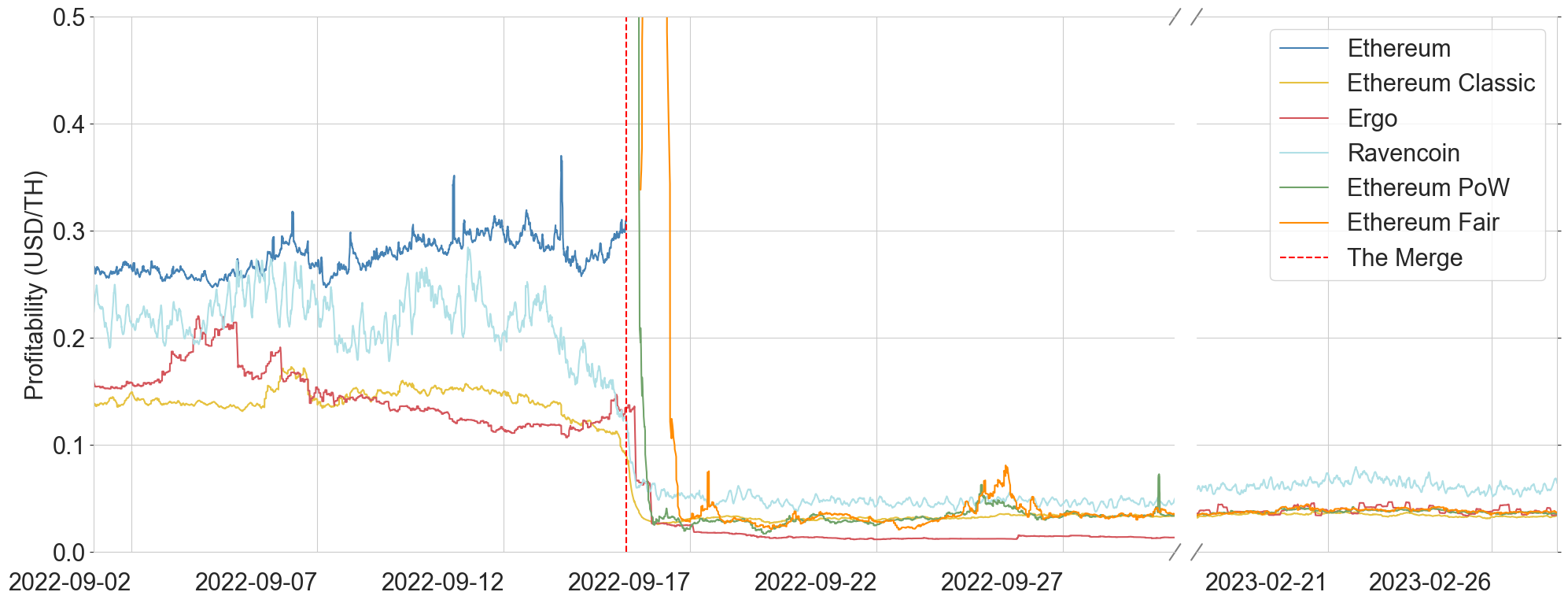}
    \end{subfigure}\\
    \caption{Expected US dollar reward per Tera-Ethash of computation.}
    \label{fig:profitability}
\end{figure}

Figure \ref{fig:profitability} reveals that a large decrease in profitability across all coins occurred after The Merge. Interestingly, the initial peaks in profitability for Ethereum Fair and Ethereum PoW are due to very few entities mining at an extremely low difficulty for 2 and 25 hours respectively. Conversely, Ergo suffered from its difficulty shooting up as large miners joined the network to reap profits. Due to the slowly adjusting difficulty function, this made Ergo block production excessively unprofitable for long periods. Eventually, EIP37 and the associated Hard Fork mitigated the issue \cite{ergo2022hardfork}, raising Ergo's profitability. These spotlight the instability period that follows such a drastic fork. Still, five months after The Merge, the profitability of each coins has converged to a value about 1/8th of what Ethereum's profitability used to be.

\section{Discussion}
\label{sec:discussion}

    \paragraph{Environmental Impact.} One of the big promises of moving to PoS was the 30,000x decrease in energy \cite{energyComparison}. Compared to the 1st of September 2022, on the 1st of March 2023, we see an excess hashrate of 116 THs (over a 2x increase) for the remaining systems. This increase has happened even though the profitability of mining is almost an order of magnitude lower. In other words, at least 12\% of the original mining power of Ethereum is still operating, dampening the immediate environmental benefits of Ethereum's switch to PoS.

    \begin{figure}[t]
            \centering
            \begin{subfigure}{.496\textwidth}
                \centering
                \includegraphics[width=\linewidth]{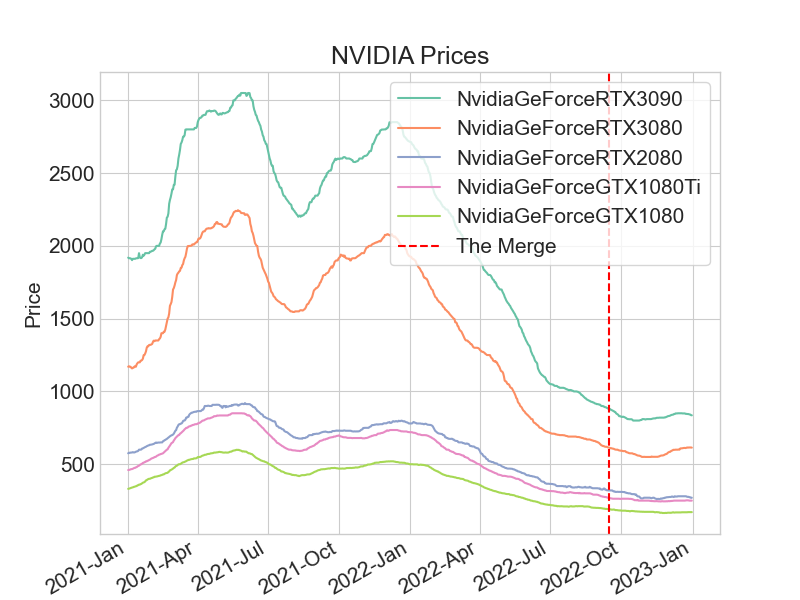}
            \end{subfigure}
            \begin{subfigure}{.496\textwidth}
                \centering
                \includegraphics[width=\linewidth]{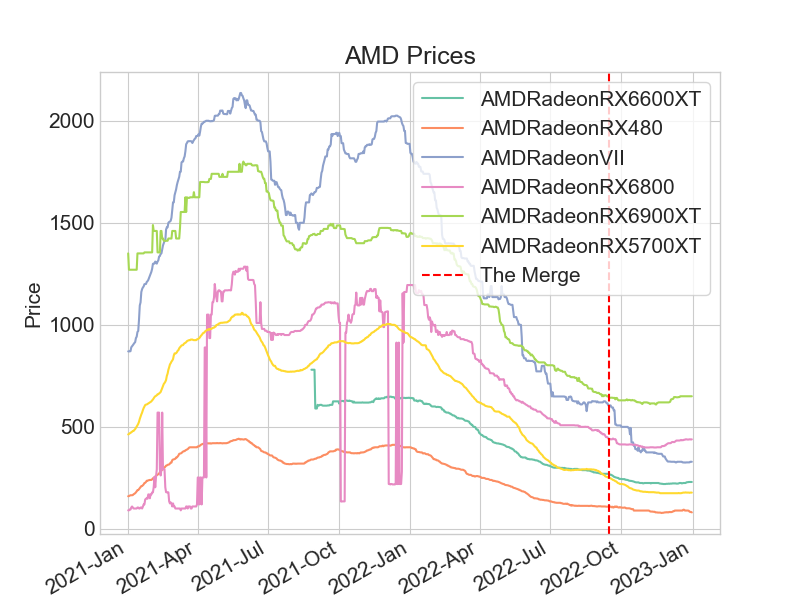}
            \end{subfigure}
            \caption{Resale value of Nvidia and AMD GPUs taken from \cite{howmuchone}.} 
            \label{fig:gpuprice}
    \end{figure}

    \paragraph{Impact on Miners.} As evidenced by the trends in resale value of the GPUs considered in this study (cf. Figure \ref{fig:gpuprice}), the demand for GPUs went down leading up to The Merge\footnote{This generally follows previously observed GPU usage cycles of miners \cite{eghbali201912}.}. Nonetheless, the reason for the excess mining power even through unprofitability can at least partially be explained through the behavior of private miners. We have found anecdotal evidence in discussion forums that some miners kept mining for i) speculative reasons, ii) to utilize heat for personal use, and iii) due to personal conviction (e.g., \cite{reddit_heat}). Others are opting to rent their GPUs on decentralized marketplaces \cite{vastai}. On the other hand, Hut8 and HIVE, two large mining companies, have been public about their strategies following The Merge, extending their ASIC mining operations and re-purposing their GPUs for AI and High-Performance Computing (HPC). Both companies, however, report major losses \cite{hut8AI,hut8mining,hivelosses,hivebitcoin,hut84}.

    \paragraph{Impact on the GPU Market.} The GPU manufacturers themselves were initially predicted to have a bleak future as late as mid-2022 \cite{nasdaq2021outlook,coindesk2022slow} and experienced year-low stock prices \cite{aim2022killing}. The situation has very quickly turned on its head thanks to the recent disruption in AI and its equally insatiable hunger for GPUs \cite{ft2023alltimehigh}. As for the blockchain space, the use of GPUs is far from over, as (zero-knowledge) proofs of knowledge are becoming a pillar of Layer-2 technologies \cite{ingonyama2023zkp}. 

\section{Related work}
\label{sec:related}
    Previous work on the Ethereum Classic fork of 2016 \cite{kiffer2017stick}, at the time the first persistent fork of its kind, studied its impact on the two systems including their profitability and mining pool distributions. They also showed a
    vulnerability caused by the shared history and code of the two systems which subsequently caused Chain-ids to be made standard in future forks. Social aspects of forks have also been studied including a case analysis of the Bitcoin Cash fork \cite{islam2019understanding}, and the ethics of planned forks \cite{kim2019ethics}. Another line of work characterises broadly the kinds of forks that exist \cite{schar2020blockchain}, including the characterization of velvet forks \cite{kiayias2020non} and their prominence and impact in Bitcoin and other networks \cite{zamyatin2018wild}. There is a rich body of work on the dynamics of mining pools, including a breakdown of the market share of pools in Ethereum \cite{kiffer2021under} and Bitcoin \cite{wang2020measurement}, as well as work on the incentives for miners to join different kinds of pools \cite{qin2018research,fisch2017socially,cong2021decentralized}. While we focus on the impact of The Merge on the general PoW ecosystems, others have studied its impact solely on the Ethereum system \cite{heimbach2023ethereum,kapengut2023event,wahrstatter2023time,wahrstatter2023blockchain}.

\bibliographystyle{plain}
\bibliography{bibtex}

\end{document}